\documentclass{article}%
\usepackage[utf8]{inputenc}
\usepackage[T2A]{fontenc}
\usepackage[OT2,OT1]{fontenc}
\usepackage{anyfontsize}
\usepackage{lmodern}
\usepackage{graphicx} 
\usepackage{amsmath}
\usepackage{authblk}
\usepackage[a4paper, margin = 1.3in]{geometry}
\usepackage{float}
\usepackage{cite}
\usepackage{changepage}
\usepackage{float}
\usepackage[caption=false]{subfig}
\usepackage[font=scriptsize]{caption}
\usepackage[colorlinks=true, allcolors=blue]{hyperref}

\usepackage{graphicx}%
\usepackage{multirow}%
\usepackage{amsmath,amssymb,amsfonts}%
\usepackage{amsthm}%
\usepackage{mathrsfs}%
\usepackage[title]{appendix}%
\usepackage{xcolor}%
\usepackage[most]{tcolorbox}
\usepackage{textcomp}%
\usepackage{manyfoot}%
\usepackage{upquote}
\usepackage{booktabs}%
\usepackage{algorithm}%
\usepackage{algorithmicx}%
\usepackage{algpseudocode}%
\usepackage{listings}%
\usepackage{graphics}
\usepackage{subfig}
\usepackage{csquotes}
\usepackage{dirtytalk}
\usepackage{xr}

\usepackage[acronym, nogroupskip]{glossaries}
\makeglossaries
\newacronym{llm}{LLM}{Large Language Model}
\newacronym{ai}{AI}{Artificial Intelligence}
\newacronym{od}{OD}{Opinion Dynamics}
\newacronym{abm}{ABM}{Agent Based Model}
\newacronym{lod}{LODAS}{Language-Driven Opinion Dynamics Model for Agent-Based Simulations}

\usepackage{setspace}
\usepackage{setspace}
\title{Language-Driven Opinion Dynamics in Agent-Based Simulations with LLMs}

\author[1,2]{Erica Cau}
\author[1,2]{Valentina Pansanella}
\author[1]{Dino Pedreschi}
\author[2]{Giulio Rossetti}

\affil[1]{Computer Science department, University of Pisa, Italy}
\affil[ ]{erica.cau@phd.unipi.it, dino.pedreschi@unipi.it}
\affil[2]{Institute of Information Science and Technologies “A. Faedo” (ISTI), National Research Council (CNR), Pisa, Italy}
\affil[ ]{{ \{valentina.pansanella, giulio.rossetti\}@isti.cnr.it}}

\date{}

\begin{document}
\maketitle
\abstract{Understanding how opinions evolve is crucial for addressing issues such as polarization, radicalization and consensus in social systems. 
While much research has focused on identifying factors influencing opinion change, the role of language and argumentative fallacies remains underexplored. 
This paper aims to fill this gap by investigating how language -- along with social dynamics -- influences opinion evolution through \acrshort{lod}, a \acrlong{lod}. 
The model simulates debates around the \say{Ship of Theseus} paradox,  in which agents with discrete opinions interact with each other and evolve their opinions by accepting, rejecting, or ignoring the arguments presented. We study three different scenarios: balanced, polarized, and unbalanced opinion distributions. 
Agreeableness and sycophancy emerge as two main characteristics of \acrshort{llm} agents, and consensus around the presented statement emerges almost in any setting. Moreover, such AI agents are often producers of fallacious arguments in the attempt of persuading their peers and -- for their complacency -- they are also highly influenced by arguments built on logical fallacies. These results highlight the potential of this framework not only for simulating social dynamics but also for exploring from another perspective biases and shortcomings of \acrshort{llm}s, which may impact their interactions with humans.}

\bigskip


\maketitle

\section{Introduction}\label{sec:introduction}
\begin{displayquote}
For the logical question of things that grow; one side holding that the ship remained the same, and the other contending that it was not the same.
\flushright{\textit{Plutarch, Life of Theseus 23.1}}
\end{displayquote}
In its original formulation, the \say{Ship of Theseus} paradox concerns a debate over whether or not a ship that had all its components replaced one by one would remain the same.
Consider engaging in a discourse regarding this paradox within the context of a philosophy class, an online Reddit community, or during a dinner gathering with friends.
Everyone will reason on the paradox and try to convince others of their stance. 
Convincing arguments can be proposed in favour or against this statement. 
Ultimately, everyone will leave the debate with their own opinion or no opinion at all. 
Regardless of the context in which the debate takes place, one thing does not change: the means through which we will try to convince our peers, or they will convince us, is \textit{language}. 
When a speaker intentionally uses language to convey a specific purpose, they exert an illocutionary force that can influence the listener's perspective, leading to a common understanding or increased division. Therefore, we must consider how language shapes the development of opinions.

\smallskip
\noindent
The development of individual and public opinions has long been a focus of psychologists and sociologists, and more recently, it has been extensively explored in computational social science \cite{conte2012manifesto,tucker2023computational} and sociophysics \cite{hobbes1651leviathan, comte1855positive}. This research acknowledges the complexity of \acrfull{od}, where multiple interacting factors lead to emergent behaviours such as consensus \cite{li2013consensus}, polarization \cite{biondi2023dynamics}, and radicalization \cite{ramos2015how}, often difficult to predict. 
Understanding the drivers of opinion change and going beyond mere observation of opinion patterns remains a complex issue.
One common approach to tackle this issue is through models of \acrshort{od}, which aim to explain how opinions evolve via social interactions \cite{castellano2009statistical}. These models simplify real-world phenomena, enabling the exploration of various what-if scenarios. They generally simulate a population of individuals and their interactions, with processes often governed by simple rules that reflect empirically observed behaviours, such as the repeated averaging of opinions with neighbours \cite{degroot1974reaching, friedkin1986formal}. Recent models also incorporate the backfire effect \cite{chen2021opinion, monti2020learning}, where individuals become more entrenched in their opinions when confronted with contradictory information \cite{nyhan2010when}.
Opinion evolution is driven by factors rooted in socio-psychological theories, such as cognitive biases \cite{allahverdyan2014opinion}, as well as external forces like peer pressure \cite{liu2021modeling}, algorithmic biases \cite{sirbu2019algorithmic}, and mass media \cite{pansanella2023mass}. While these models provide simplified representations of societal dynamics and help stakeholders understand social behaviours, they often overlook important complexities. For example, they typically map opinions and messages to numerical values and rely on rule-based agents, which limits their ability to capture the nuances of human behaviour and the complex relationships between agents' characteristics, such as demographics and personality traits.

\smallskip 
\noindent
To overcome such limitations, we propose a novel framework exploiting \acrlong{llm}s (\acrshort{llm}s) capabilities to create an \acrfull{abm} that allows for the study of the interplay between language and opinion change in the long term.
The relationship between language and opinion change has been underexplored. Monti et al. \cite{monti2022language} is a prominent exception, highlighting the role of knowledge, similarity, and trust in a social media case study. 
Their findings challenge simplistic \acrshort{od} models, emphasizing the need for more complex analysis.
LLMs have revolutionized language-related studies, enabling more realistic social simulations. 
Park et al. \cite{Park2022, park2024generativeagentsimulations1000} introduced \acrshort{llm} agents as \textit{social simulacra}, capable of simulating personalities and social behaviours. 
Claims about \acrshort{llm}s possessing Theory of Mind (ToM) \cite{Premack1978} remain debated: while Kosinski \cite{kosinski2023theory} and others \cite{Street2024LLMsachieve, li2023theory} suggest they exhibit emergent ToM abilities, critics \cite{Ullman2023large, sap-etal-2022-neural, shapira2023well} highlight their inconsistencies in ToM tasks and lack of genuine social intelligence. 
Nevertheless, even a simulated ToM may enhance \acrshort{od} models by enabling agents to consider interlocutors' mental states.
\acrshort{llm}-driven populations display emergent behaviours akin to human societies, such as scale-free networks \cite{de2023emergence} and information diffusion \cite{gao2023s}. 
In opinion evolution, \acrshort{llm} agents replicate echo chambers \cite{mou2024unveiling}, polarization \cite{wang2024decoding}, and confirmation bias effects \cite{chuang2023simulating}. 
While \acrshort{llm}s can generate persuasive arguments \cite{breum2023persuasive} aligned with psycho-linguistic theories \cite{monti2022language}, they are less convincing than humans \cite{flamino2024limits} and exhibit biases toward scientific accuracy \cite{chuang2023simulating}, politeness \cite{priya2024computational}, and platform-specific discourse styles \cite{tornberg2023simulating}. 
Despite these biases, \acrshort{llm}-based agents have successfully reproduced experimental results in psychology and linguistics \cite{aher2023using}, making them valuable tools for \textit{in silico} social experiments.

\begin{figure}[h]
    \centering
    \includegraphics[width=\linewidth]{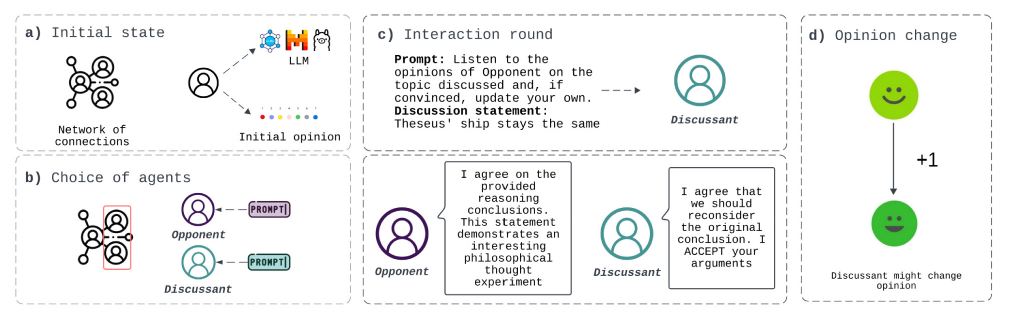}
    \caption{\textbf{Graphical schema of \acrshort{lod}.} The \acrshort{llm} agents population is initialized as a network; each agent is an \acrshort{llm} instance with an initial opinion in the range [0, 6] (a). At each iteration, two agents are randomly chosen and prompted to act as \textit{Opponent} and \textit{Discussant} (b). The \textit{Discussant} is prompted to listen to the opinion of the \textit{Opponent} around the discussion statement and may then accept, reject or ignore such opinion (c) and update their current one accordingly by ±1 (d).
    }
    \label{fig:framework}
\end{figure}

\smallskip
\noindent
In this paper, we explore the interplay between language and social interactions in shaping \acrshort{od} through \acrshort{lod} a \acrlong{lod}. 
A schematic representation of \acrshort{lod} is provided in Figure \ref{fig:framework}. 
As shown in Figure \ref{fig:framework}(a), \acrshort{llm} agents (instances either of Mistral or Llama3 models) hold one of seven possible opinions, evolving through social interactions via ±1 updates. 
Using a 7-point scale follows established Likert-scale \cite{likert1932likert} methodologies in psychological research for measuring subjective constructs.
We simulate three distinct scenarios: 
(i) a baseline scenario with a uniform opinion distribution; 
(ii) a polarized scenario, where opinions are bimodally distributed between positive and negative stances with no neutral positions; 
and (iii) an unbalanced scenario, where most agents initially hold an extremely negative stance. 
Throughout the simulations, two agents are selected at random (see Figure \ref{fig:framework}(b)) to engage in discussion (see Figure \ref{fig:framework}(c)), where the \textit{Opponent agent} (\textit{Opponent}, from now on) attempts to persuade the \textit{Discussant agent} (\textit{Discussant}, from now on), who may then update their opinion on the Ship of Theseus paradox.  
This topic was chosen to minimize controversy and prevent convergence toward a predefined ground truth, a phenomenon documented in prior studies \cite{breum2023persuasive, chuang2023simulating, chuang-etal-2024-simulating}. 
To assess the impact of linguistic framing, we start the discussion with one of two formulations: (i) a positive direction (\say{The ship remains the same}) and (ii) a negative direction (\say{The ship becomes different}). This choice follows prior research \cite{chuang2023simulating} demonstrating how initial statement framing (\say{Global warming is/is not a hoax}) may influence opinion evolution.

\smallskip
\noindent 
This study aims to advance \acrlong{od} and social simulations by leveraging Large Language Models (LLMs). Traditional models rely on mechanistic assumptions rarely validated in real-world settings, limiting their applicability. Instead, we explore whether \acrshort{llm} agents, operating without predefined update rules and guided by the Theory of Mind (ToM) hypothesis, can exhibit realistic individual behaviour and emergent collective dynamics. 
Unlike classical models, \acrshort{llm} agents engage in natural language interactions, allowing us to investigate the interplay between language and opinion change. Specifically, we examine how they employ and propagate logical fallacies and assess their role in persuasion—an aspect understudied in prior research, primarily focused on fallacy detection. A notable exception is Breum et al. (2023) \cite{breum2023persuasive}, who analyzed \acrshort{llm}-driven persuasion, showing that trust, status, and knowledge influence stance shifts. However, their study focused on one-shot interactions, while we examine how \acrshort{llm}s adapt arguments and leverage fallacies over time. Payandeh et al. (2023) \cite{Payandeh2023} provide the first systematic analysis of \acrshort{llm}s' susceptibility to fallacious reasoning in debates. They find that GPT-4 agrees with flawed arguments 67\% of the time, significantly more than logically sound ones. Building on this, we investigate how \acrshort{llm}s not only process but also generate fallacies in multi-agent interactions, shedding light on their role in long-term opinion evolution.

\smallskip
\noindent 
The remainder of this paper is organized as follows. 
In Section \ref{sec:results}, we examine the outcomes of our simulations across different initial conditions and scenarios, analyzing, on the one hand, opinion trends, acceptance rates, and, on the other, the linguistic patterns in agent interactions, assessing the role of logical fallacies in shaping opinion change. 
Section \ref{sec:methodology} details the simulation framework and experimental design. 
In Section \ref{sec:discussion}, we discuss our findings, highlighting a key emergent behaviour: a strong tendency toward consensus driven by \acrshort{llm}s' inclination to agree with presented statements. 
We also highlight the prevalence of fallacies in \acrshort{llm}-generated discourse and their impact on persuasion. 
Additionally, in Section \ref{sec:conclusion}, we outline key takeaways, study limitations, and directions for future research. 
Additional figures and analyses are provided in the Supplementary Materials.

\section{Results}
\label{sec:results}
This work extends the modelling of \acrlong{od} using \acrshort{llm} agents to explore whether and which emergent behaviours arise without explicit opinion modification rules. Additionally, it examines the linguistic features of the debates, linking them to specific agent roles and behaviours.
To this end, we defined a framework in which a networked population of \acrshort{llm} agents discusses a given topic, updating their opinions according to tunable behavioural rules.
Our simulations considered a population of 140 \acrshort{llm} agents. 
We assumed a mean-field context (i.e., all agents can interact with all other agents without any social restrictions), a commonly used starting point to identify potential emerging behaviours from the opinion evolution process.
Each agent is an \acrshort{llm} instance, 
holding a discrete opinion in the interval [0, 6], where 0 means \textit{strongly disagree} and 6 \textit{strongly agree} with a given statement.  
Agents -- as in many classical \acrshort{od} models -- interact with each other at discrete time intervals in a pairwise fashion: at each time step, an interacting pair is chosen at random among the connected agents; in this way, in each interaction, we can assign each agent one of two roles, respectively \textit{Opponent} and \textit{Discussant}.

In the present work, we assigned as a discussion topic the paradox of the \textit{Ship of Theseus}, a thought experiment on the concept of identity first recorded in Plutarch's works.
The rationale behind the paradox is the following: if all the parts of the ship are replaced over a long period, is the resulting ship the same ship it was at the beginning? 
This dilemma was chosen because there is no scientific truth. In this way, we avoid \acrshort{llm}s converging toward what they know to be scientifically valid and limit their bias toward immediate adherence to positive opinions.
We designed our model to pose this \say{dilemma} in two different ways: (i) \say{the boat is the same}, and (ii) \say{the boat is not the same}.
Different initial conditions were analyzed. In the balanced scenario, opinions were uniformly distributed (20 agents assigned to each opinion), (ii) in the polarized scenario we removed the neutral and moderate opinions, leaving only the extreme ones, for a total of 69 agents with a strongly agreeing position and 72 extremely disagreeing with the statement, and (iii) in the unbalanced scenario a majority in the population holds an extreme (negative) opinion, to account for the tendency of \acrshort{llm}s towards agreement and against conflicts. 
We leveraged Mistral-7B Instruct \cite{jiang2023mistral7b} (Mistral from now on) and Llama-3-8B \cite{dubey2024llama} (Llama from now on) to compare different open state-of-the-art \acrshort{llm}s. 
By varying the direction of the dilemma, the \acrshort{llm}, and the initial distribution of opinions, we created 12 distinct settings.
From our simulations, we obtained opinion evolution data and related textual data, allowing us to relate language and opinion change.

In the following, we will discuss the differences and similarities across our settings based on these two main dimensions, identifying which findings are generalizable or specific to the language model.

\subsection*{Balanced scenario}
In the first setting, we analyzed a balanced scenario in which agents' initial opinions were uniformly distributed, with each opinion held by 20 agents.
This choice was made for several reasons: first of all, a uniform initial distribution is normally considered the starting condition for evaluating the effects of an opinion evolution model in classical \acrshort{od}, moreover, this creates a \say{neutral baseline} against which we can compare other distinct initial conditions, e.g., polarized or unbalanced, and account for the impact of the initial state on the final one, which is known to be a determining factor in many social dynamics models. 

\subsubsection*{Opinion dynamics}

\begin{figure}[t!]
    \centering
    \includegraphics[width=\textwidth]{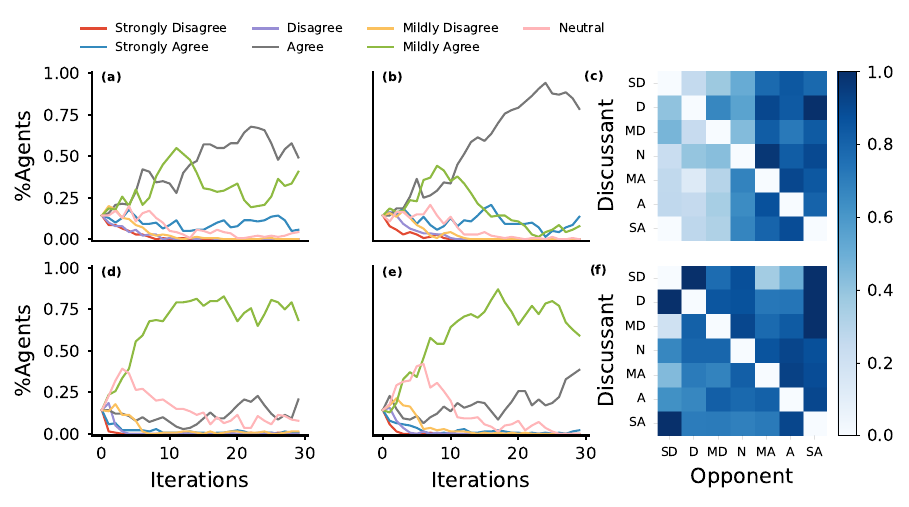}
    \caption{\textbf{Balanced scenario - Mistral and Llama-agents opinion trends and acceptance rates.} Mistral (a)-(b) and Llama (d)-(e) opinion trends for the positive (a)-(d) and negative (b)-(e) statements. Trends are represented for Strongly Disagree (red), Disagree (lilac), Mildly Disagree (yellow), Neutral (pink), Mildly Agree (green), Agree (grey) and Strongly Agree (blue) opinions. Matrices represent acceptance rates, i.e., the probability that the Discussant accepts (as in, moves towards) the opinion of the Opponent for (c) Mistral and (f) Llama agents for the positive statement.}
    \label{fig:balanced}
\end{figure}

We started by analyzing the opinion evolution patterns in the balanced scenario.
Figure \ref{fig:balanced} shows the evolution of the percentage of agents for each discrete opinion over time (with a maximum number of time steps set to 30). 
Regardless of the \acrshort{llm} model employed and regardless of the direction of the statement (\say{positive} or \say{negative}), we can see that the population reaches an agreement around the presented statement. 
More specifically, the dynamics shown by the trend lines in Figure \ref{fig:balanced}(a)-(b) and \ref{fig:balanced}(d)-(e), which is qualitatively similar across different runs of the same experiment 
(see Supplementary Figures S1-S2), shows how negative and neutral opinions tend to quickly disappear in the first timesteps; on the opposite side, positive opinions grow, as we can see from the growth of the \textit{mildly agree} opinion (green) from the beginning of the dynamics. 
Interestingly, after around 20 iterations, agents who agree outgrow all the others, and, in the end, either all the population reaches a consensus around the \textit{agree} position or at least a majority of agents reaches the \textit{agree} position, while a minority remains holding the \textit{mildly agree} position. 

Comparing the results across the positive and the negative statement, what emerges is that with the positive statement (Figure \ref{fig:balanced}(a) and Figure \ref{fig:balanced}(d)), there is a positive percentage of agents holding a \textit{strongly agree} opinion, while the majority still holds the \textit{agree} opinion. 
When presented with the negative statement (Figure \ref{fig:balanced}(b) and Figure \ref{fig:balanced}(e)), only the \textit{strongly agree} opinion survives.

If we compare Mistral agents to Llama agents, we can see that the populations of Mistral agents have a stronger tendency towards agreement, with negative opinions disappearing from the population faster than Llama agents populations, when presented with the negative statement. 
Instead, with the positive one, there are still no major differences between the two models. 
However, Mistral agents show a more unstable dynamic, with agents switching between strongly agree, agree, and mildly agree, and simulations ending after 100 iterations with the presence of all three opinions in the final state.

\begin{tcolorbox}[colback=gray!5!white,colframe=gray!75!black,title=Highlight]
LLM populations tend to collectively agree with the given statement.
\end{tcolorbox}

These trends highlight a tendency towards agreement that also emerges from analysis of the acceptance rate matrices in Figures \ref{fig:balanced}(c) and \ref{fig:balanced}(f). 
Each cell of the matrix represents the acceptance rate of the \textit{Opponent}'s opinion by the \textit{Discussant}, i.e. the percentage of interactions in which the \textit{Discussant}'s opinion changes by a unit amount (±1) in the direction of the \textit{Opponent}'s opinion.

Changing the way the proposition is presented has no effect on the probability of accepting the \textit{Opponent}'s opinion. 
Since the probabilities are very similar, we only show matrices for the positive statement. 
Matrices for the negative statement can be found in the supplementary information (see supplementary figures S7-S8).
However, if we compare Figure \ref{fig:balanced}(c) with Figure \ref{fig:balanced}(f), we see that the two different language models generate agents that behave differently. 
For Mistral agents (Figure \ref{fig:balanced}(c)), the probability of acceptance increases with the opinion of the \textit{Opponent}: the higher the opinion, the higher the probability of acceptance.  
Regardless of their current stance, agents are generally reluctant to accept or be persuaded by disagreeing opinions, even mild ones.
Instead, Llama agents (Figure \ref{fig:balanced}(f)) tend to accept the \textit{Opponent}'s opinion with a high probability across all conditions. 
Although agreeing agents are generally more likely to accept opinions that align with their own, they still exhibit a 100\% acceptance probability for strongly disagreeing opinions, for instance. 
As noted earlier, this behaviour holds true for discussions surrounding both positive and negative statements. While small differences appear, the overall pattern -- or lack thereof -- remains qualitatively consistent. 

\begin{tcolorbox}[colback=gray!5!white,colframe=gray!75!black,title=Highlight]
The acceptance rate of different opinions varies across \acrshort{llm}s. Mistral agents tend to accept agreeing opinions at high rates and negative opinions at low rates. Llama agents tend to accept any opinion at high rates.
\end{tcolorbox}

\subsubsection*{Logical fallacies analysis} 
Given the two distinct roles we assigned to our \acrshort{llm} agents -- \textit{Opponents} generating persuasive statements and \textit{Discussants} justifying their opinions -- we examined how a different framing of the discussion statement impacts the linguistic behaviour in general and, more specifically, the emergence of logical fallacies in the generated arguments and their interplay with the persuasiveness of the agents.

\noindent\textbf{Language variability}\\
\textit{Opponents} generated fewer statements than the \textit{Discussants}, as a set of utterances were repeated across multiple iterations with variations attributable to LLMs (Table \ref{tab:linguisticfeatures}). 

Specifically, in the balanced scenario -- and similarly in all the others -- Llama generated only an average of 50-55\% unique statements. 
Mistral exhibited greater variability, with an average of 67\% of unique statements in the balanced scenario and a drop to 22.81\% in the polarized scenario.
In the unbalanced scenario, the number of unique statement marginally increased to 34\% in the \say{same boat} while the 42\% characterizing the \say{different boat} scenario was similar to the 45\% registered by Llama in the same setting.

Conversely, the \textit{Discussants} did not suffer this issue as much, showing more significant variability and adaptability when motivating their stance in most cases.
Exceptions relate the initial setup -- polarized or unbalanced -- and the framing of the statement, which caused a drop in both Llama and Mistral.



\begin{table}[ht] 
\small
\setlength{\tabcolsep}{1.5pt}
\begin{tabular}{|l|cc|cc|cc|cc|cc|cc|}
\hline
 & \multicolumn{4}{c|}{Balanced} & \multicolumn{4}{c|}{Polarized} & \multicolumn{4}{c|}{Unbalanced} \\ \hline
 & \multicolumn{2}{c|}{Llama} & \multicolumn{2}{c|}{Mistral} & \multicolumn{2}{c|}{Llama} & \multicolumn{2}{c|}{Mistral} & \multicolumn{2}{c|}{Llama} & \multicolumn{2}{c|}{Mistral} \\ \hline
 & S & D & S & D & S & D & S & D & S & D & S & D \\ \hline
N. Sentences (\textit{O}) & 55.37 & 56.43 & 71.51 & 65.76 & 50.44 & 29.74 & 21.25 & 22.81 & 66.07 & 45.85 & 34.06 & 42.35 \\ \hline
N. Sentences (\textit{D}) & 92.70 & 92.57 & 91.82 & 83.72 & 40.44 & 43.99 & 26.78 & 31.98 & 95.40 & 41.60 & 36.81 & 60.29 \\ \hline
\end{tabular}
\caption{Percentage of unique sentences over the total, produced by \textit{Discussants} (\textit{D}) and \textit{Opponents} (\textit{O}). \textit{S} and \textit{D} stand for the two configurations with the ``the boat is the same" (S) and ``the boat is different" (D) discussion statements.} \label{tab:linguisticfeatures}  
\end{table}

\begin{figure}
    \centering
    \includegraphics[width=\textwidth]{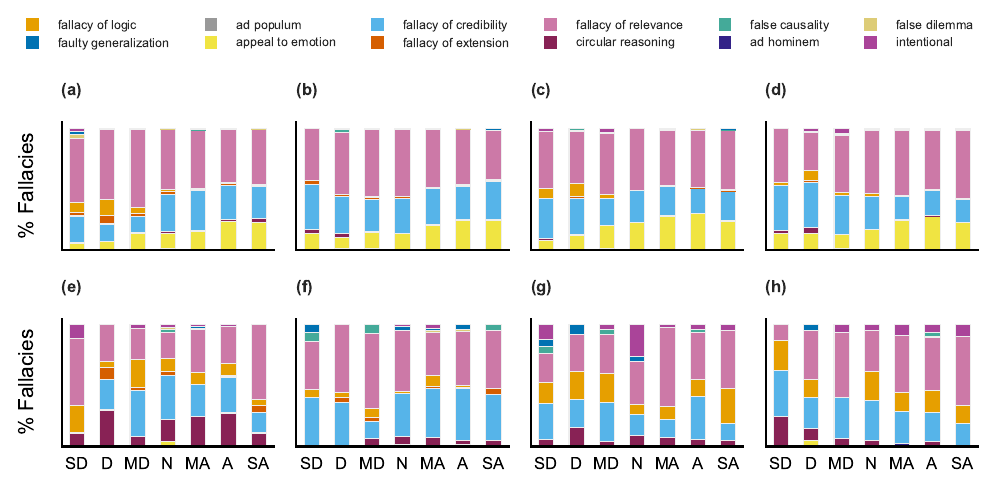}
    \caption{\textbf{Logical fallacies distribution in the balanced scenario.} Percentage of logical fallacies by opinion, shown for Llama and Mistral.
    Panels (a)-(b) refers to Llama agents discussing the ``same boat" (a) and ``different boat" scenario (b). Panels (c)-(d) refers to Mistral discussing the same (c) and different boat (d) framing. 
    Panels below refer to Llama (e)-(f) and Mistral (g)-(h) \textit{Discussants}). 
    Fallacies of relevance (pink) and fallacies of credibility (light blue) are the most common overall across \textit{Opponents}, followed by appeal to emotion (yellow). \textit{Discussants} of both LLMs adds the circular reasoning fallacy (plum).}
    \label{fig:fallaciesBalanced}
\end{figure}

Starting from these preliminary insights on the interplay between initial opinion distribution and opinion shifting, we took advantage of the opportunity offered by \acrshort{llm}-based agents to study the role of language in opinion change, identifying emergent behaviours. 
Opinion analysis revealed a strong tendency toward agreement, with agents consistently converging around the \textit{agree} or \textit{strongly agree} positions and frequently over-accepting others' opinions. This alignment may be influenced by specific linguistic patterns or the incorporation of linguistic biases into generated responses.
Therefore, we analyzed the presence of logical fallacies to quantify their role in the persuasive strategies and opinion changes in the performed simulations.
Both Llama and Mistral \textit{Opponents} (Figure \ref{fig:fallaciesBalanced} (a)-(d) demonstrate a trend toward convincing through fallacies of relevance, logical flaws occurring when a statement lacks logical coherence as the premises do not adequately support the conclusion.
In addition, \textit{Opponents} also produced sentences containing fallacies of credibility and leveraged the appeal to an irrational emotion without logical reasoning behind it, producing a fallacy of \textit{appeal to emotion}. 
Figure \ref{fig:fallaciesBalanced} shows more in detail the distribution of fallacies for each model and debated statement.
Fallacies are consistent across \acrshort{llm}s and statement directions (Figure \ref{fig:fallaciesBalanced} (a)-(b) for Llama and Figure \ref{fig:fallaciesBalanced} (c)-(d) for Mistral \textit{Opponents}).
Fallacies of relevance were found in all four cases, appearing in almost half of the statements, with the exception of the \textit{strongly disagree} and \textit{disagree} cases in Mistral with the negative statement.
Another noteworthy fallacy that emerged was the fallacy of credibility, where agents increased the credibility of their statements by appealing to ethics or authority.
The average occurrence of this fallacy reached its maximum in Llama3 in the \say{different boat} setting (30\%, in Figure \ref{fig:fallaciesBalanced}(d)), while it was lowest at 24\% in the Llama \say{same boat} configuration (Figure \ref{fig:fallaciesBalanced}(c)).

To gain further insight into the connection between fallacies and persuasion, we investigated whether the use of fallacies by an \textit{Opponent} was associated with a change of opinion in the \textit{Discussant}.
Table \ref{tab:opinion_change_rates} shows the effects of the fallacies on the different scenarios of balanced, polarized and unbalanced opinion distribution.
While Llama3 (Table \ref{tab:opinion_change_rates}) had the highest percentage of successful persuasion of the \textit{Discussant} through fallacies, with 72\% in the case of the \say{same boat} scenario and 75\% in the \say{different boat}. 
Conversely, Mistral managed to change the opinion of the \textit{Discussant} through fallacies 45\% of the time in the \say{different boat} scenario and 49\% of the time in the \say{same boat} scenario.

\begin{tcolorbox}[colback=gray!5!white,colframe=gray!75!black,title=Highlight]
Most of the fallacious statements from both Mistral and Llama agents involve fallacies of relevance and credibility.
Llama \textit{Opponents} have a higher success rate of persuading \textit{Discussant} agents when using fallacious statements than Mistral.
\end{tcolorbox}

The situation changes considerably when we examine the text produced by the \textit{Discussants} after reading the \textit{Opponent} statement. 

Panels (e)-(h) in Figure \ref{fig:fallaciesBalanced} refer to Llama agents discussing the \say{same boat} (e) and \say{different boat} scenario (f), while panels (g)-(h) refer to Mistral symmetrically. 
In addition to the relevance and credibility fallacies seen in the \textit{Opponents}, new fallacies emerge. 

Most notably, circular reasoning -- a fallacy where conclusions rely on the premises and vice versa -- appears frequently among Llama agents discussing the \say{same boat} statement, as shown in Figure  \ref{fig:fallaciesBalanced}(a).
The prevalence of relevance and credibility fallacies among \textit{Discussants} does not seem to follow a clear pattern but varies with each scenario and input opinion (Figure \ref{fig:fallaciesBalanced}(e)-(h)).

\begin{tcolorbox}[colback=gray!5!white,colframe=gray!75!black,title=Highlight]
Answers given by \textit{Discussants} - especially by Llama in the \say{same boat} discussion - introduce circular reasoning fallacy, where the conclusion is used as the premise of the argument.
\end{tcolorbox}

\subsection*{Polarized scenario}
With agreeableness and condescension emerging from this first scenario, one could ask if the presence of \say{moderate} agents may bridge opposing views and, in the end, lead the population towards a positive consensus.
In order to examine the role of such balanced initial condition and to explore the differences with a more divided start, we divided populations into two clusters: one holding the \textit{strongly agree} opinion and one holding the \textit{strongly disagree} opinion, creating a polarized initial condition.
Debates in the \textit{public sphere} are often characterised by a pronounced dichotomy between two conflicting viewpoints without moderate stances being able to bridge between the two. 
A prominent example is the United States, where political debates have become increasingly polarized \cite{pew2023dismal}, leading to a growing divide between the Democratic and Republican parties, where each side firmly defends its position on various issues.
From a linguistic perspective, persuasive discourse in such debates often relies on fallacious reasoning, something that has been previously highlighted in other works \cite{zurloni2013fallacies, abbas2024fallacy}. 
As the agents' discourse showed logical fallacies since the balanced scenario, this kind of analysis is essential to assess if logical fallacies are present and whether they are able to influence consensus formation even with \acrshort{llm} agents.

Although our simplified scenario does not capture all the complexities of these dynamics, we use it to test the hypothesis that our simple \acrshort{lod} model can reproduce the dynamics of a bipartition of the population into clusters that hold, if not extreme, at least opposed opinions.

\subsubsection*{Opinion dynamics}
\begin{figure}[]
    \centering
    \includegraphics[width=\textwidth]{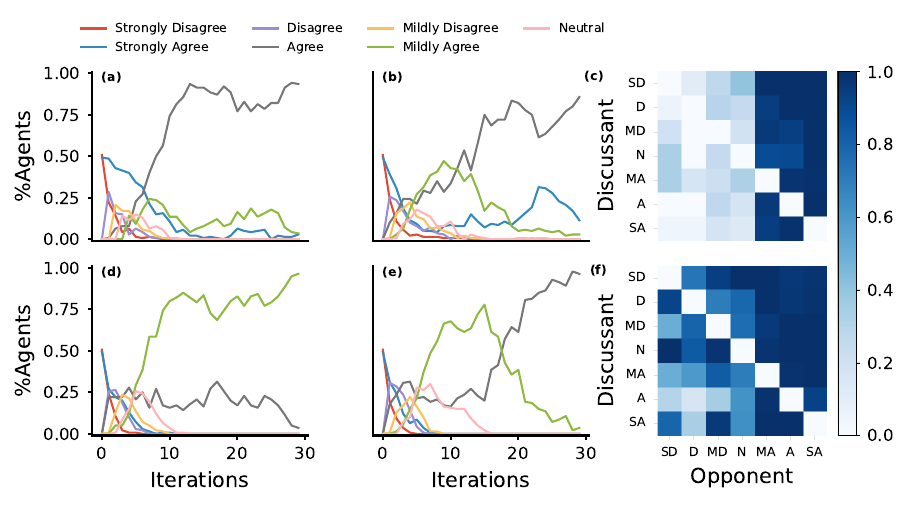}
    \caption{\textbf{Polarized scenario - Mistral and Llama-agents opinion trends and acceptance rates.} Mistral (a)-(b) and Llama (d)-(e) opinion trends for the positive (a)-(d) and negative (b)-(e) statements. Trends are represented for Strongly Disagree (red), Disagree (lilac), Mildly Disagree (yellow), Neutral (pink), Mildly Agree (green), Agree (grey) and Strongly Agree (blue) opinions. Matrices represent acceptance rates, i.e., the probability that the Discussant accepts (as in, moves towards) the opinion of the Opponent for (c) Mistral and (f) Llama agents for the positive statement.}
    \label{fig:polarized}
\end{figure}

Following the same structure as for the balanced scenario, we examine opinion evolution patterns. 
As shown in Figure \ref{fig:polarized}, the population quickly converges towards agreement irrespective of the starting polarized condition.
No matter which \acrshort{llm} model is used to enhance agents, and regardless of whether the statement is presented as positive or negative, we observe that the population reaches a consensus on the statement, albeit with varying degrees of strength in that agreement.

As we can see from \ref{fig:polarized}(a)-(b) and \ref{fig:polarized}(d)-(e) \textit{strongly disagreeing} agents (blue) quickly adopt the opinions from the other side of the spectrum, shifting towards less disagreeing and then agreeing opinions. 
Even if at a lower rate, \textit{strongly agreeing} agents (orange) start accepting a milder position and -- as we saw in the balanced scenario -- the convergence towards the \textit{agree} opinion is boosted, leading the population to quickly reach the consensus (or at least a majority) around that opinion (red line). 

Similarly to what emerged from the balanced scenario, changing the framing of the statement under discussion from positive (\say{same boat}) to negative (\say{different boat}) created a more unstable dynamic, with clusters of agreeing opinions coexisting for a longer time (Figure \ref{fig:polarized}(b) and \ref{fig:polarized}(e)). 

The difference we see in the opinion evolution trend between Mistral and Llama agents is that the latter tend to converge more closely on \textit{mildly agree} than \textit{agree} or \textit{strongly agree}, especially if the statement is discussed in its positive form (\say{the boat is the same}). 
When it is discussed in its negative form, \textit{mildly agree} resists within the population for a longer time with respect to the Mistral case; however, in the end, the population converges around the \textit{agree} opinion. 
\begin{tcolorbox}[colback=gray!5!white,colframe=gray!75!black,title=Highlight]
Polarized initial conditions do not lead to the coexistence of opposite opinions or convergence towards neutral opinions. 
LLM agents always converge towards a single opinion and \textit{agree} on the presented statement. 
\end{tcolorbox}

Additional figures on the opinion prevalence trends in the polarized scenario can be found in the Supplementary Information (see Supplementary Figures S3-S4). 

\smallskip
Surprisingly, the matrices in Figure \ref{fig:polarized} reveal that starting from a polarized initial condition enhances the acceptance rate of positive opinions. This effect is more pronounced in Mistral agents than Llama agents.
In the former, the acceptance rate of agreeing opinions, regardless of the agent's initial stance, is always around 1.0.
In contrast, Llama agents, as previously emerged from the balanced scenario, tend to have higher acceptance rates for all the opinions presented (agreeing or disagreeing with the statement).
However, in this case, while the acceptance rate of the disagreeing opinions varies from 0.28 to 0.98, the acceptance rate of agreeing opinions is almost always around 1.0. 
These patterns can be identified when discussing either the \say{positive} form or the \say{negative} form of the statement, with just a small variation of the rate values. 

\begin{tcolorbox}[colback=gray!5!white,colframe=gray!75!black,title=Highlight]
The acceptance rate of different opinions varies across \acrshort{llm}s, and polarized initial conditions enhance the acceptance rate of agreeing opinions with respect to a balanced initial scenario.
\end{tcolorbox}

Additional matrices can be found in the Supplementary Information (see Supplementary Figures S9-S10).

\subsubsection*{Logical fallacies analysis} 
The number of unique utterances produced by \textit{Opponents} and \textit{Discussants} was significantly lower in the polarized setup than in the balanced one. 
This was particularly evident when using Mistral agents: about the 22\% of statements \textit{Opponents} produced to persuade \textit{Discussants} were unique in both scenarios (Table \ref{tab:linguisticfeatures}), while in all other iterations these statements were repeated.

Conversely, half of the utterances used by Llama agents when discussing the \say{same boat} prompt were utilized once.
However, in the negative statement, this originality decreased, with only 30\% of agents generating unique statements.

\begin{figure}[!h]
    \centering
    \includegraphics[width=\textwidth]{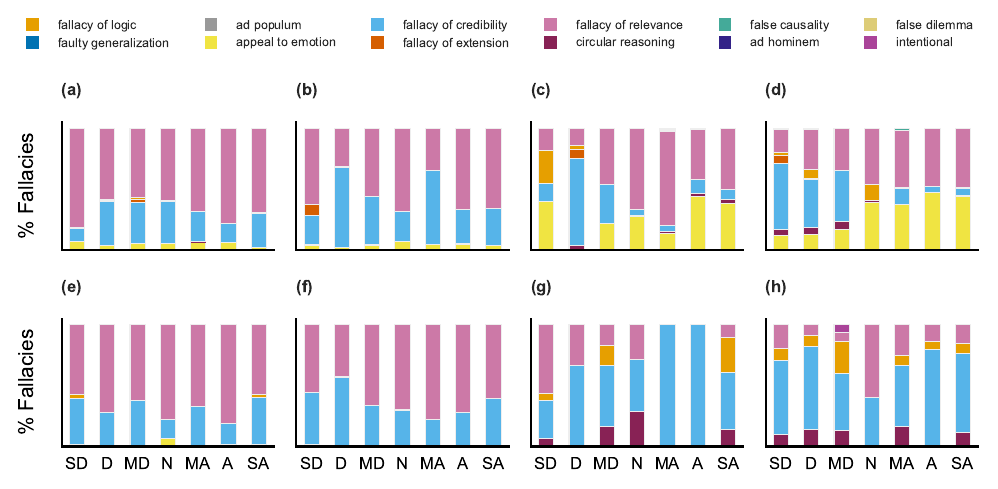}
    \caption{\textbf{Logical fallacies frequency by opinion within polarized initial conditions.}
    Percentage of logical fallacies produced by the \textit{Opponents} (on the top) and \textit{Discussants} (on the bottom) in polarized initial conditions. Panels (a) refers to Llama \textit{Opponents} discussing the ``same boat" statement, while (b) refers to the ``different boat" statement. Panels (c)-(d) similarly refer to Mistral agents. On the bottom, Llama \textit{Discussants} (e)-(f) and Mistral \textit{Discussants} (g)-(h).}
\label{fig:fallaciesPolarized}
\end{figure}

In general, agents in the polarized setup showed a stark tendency to produce flawed reasoning and arguments when discussing or persuading another agent, as was also observed in the balanced scenario. 
With both initial opinion distributions, Llama3 agents tend to produce more fallacious arguments when arguing with \textit{Opponents}, while Mistral agents produce more reasoned arguments to support and discuss the claim.
Llama agents mainly discuss through fallacies of relevance and fallacies of credibility (see Figure \ref{fig:fallaciesPolarized}(a) for the \say{same boat} and (b) for the \say{different boat} statement).
This tendency is particularly noticeable when Llama agents hold the \textit{strongly disagree} opinion discussed the \say{same boat} statement (Figure \ref{fig:fallaciesPolarized}(a)), while fallacies of credibility are primarily found in agents with a \textit{disagree} opinion or who \textit{mildly agreed} with the \say{different boat} starting prompt, as described by Figure \ref{fig:fallaciesPolarized}(b). 

However, Mistral agents produced a greater variety of fallacies.
In both positive and negative framing of the statement (Figure \ref{fig:fallaciesPolarized}(c)-(d)), agents attempted to persuade their \textit{Discussants} by appealing to emotion with a 30\% probability, which is higher than in the balanced scenario.
This fallacy was only absent in the case of agents who disagreed with the \say{same ship} statement (Figure \ref{fig:fallaciesPolarized}(c)), where sentences with credibility fallacies are prevalent.
Conversely, agents expressing a \textit{mildly agreeing} opinion exhibited the least persuasion through emotion, though their sentences predominantly exhibited fallacies of relevance. 
Mistral \textit{Opponents} behave differently according to the polarity of their opinion, esp. 
As shown in Figure \ref{fig:fallaciesPolarized}(d), agents in the \textit{disagreeing} range, i.e., from \textit{strongly} to \textit{mildly disagree}, mainly produced fallacies of credibility, whereas agents from the \textit{neutral} to \textit{fully agree} opinion range, mainly produced fallacies of relevance or appealed to the emotions of the \textit{Discussant}.

After assessing the presence of fallacies and their typology, we moved on to the analysis of the persuasiveness of the \textit{Opponents} with respect to the \textit{Discussants}' positions.
Table \ref{tab:opinion_change_rates} displays the percentage of opinion changes due to fallacious statements. 
The polarized scenario exhibited the most significant opinion changes in the \textit{Discussant} as a result of fallacies in the \textit{Opponent}'s statements. This effect was particularly notable for Llama agents, both in the \say{positive} and \say{negative} settings, as well as for Mistral agents in the \say{same boat} scenario, where 68\% of opinion changes followed a fallacious statement. 
In contrast, in the \say{different boat} scenario, Mistral agents showed fewer opinion changes after fallacious statements, although the rate remained above 50\%.

When examining the \textit{Discussants}' answers, a clear difference between \acrshort{llm}s emerged.
Llama-agents, in \say{same boat} (Figure \ref{fig:fallaciesPolarized}(e)) and \say{different boat} (Figure \ref{fig:fallaciesPolarized}(f)) settings, mainly used fallacies of relevance and fallacies of credibility, with the relative frequency varying with the opinion of each agent.
In almost all cases, fallacies of relevance were the most common, except for agents \textit{disagreeing} with the \say{different boat} statement (Figure \ref{fig:fallaciesPolarized}(f)), where fallacies of credibility were more common.
In contrast, Mistral's agents (Figure \ref{fig:fallaciesPolarized}(g)-(h)) were particularly prone to using credibility fallacies in almost all cases.
Among the exceptions, agents with a \textit{strongly disagree} opinion in Figure \ref{fig:fallaciesPolarized}(g) and neutral agents in Figure \ref{fig:fallaciesPolarized}(h) produced almost 60\% of fallacious sentences employing fallacies of relevance. 

\begin{tcolorbox}[colback=gray!5!white,colframe=gray!75!black,title=Highlight]
When in a polarized scenario, Llama agents mostly produced sentences with fallacies of relevance and credibility. Mistral agents tend to persuade with an appeal to the interlocutor's emotions.
\end{tcolorbox}

\subsection*{Unbalanced scenario}
We decided to explore an initial unbalanced distribution of opinions, with approximately 60\% of agents expressing \textit{strongly disagree} with the given statement. 
This choice was made to assess the propensity of \acrshort{llm} agents to converge towards the more positive end of the scale (slightly agree, agree, and \textit{strongly agree}).
The population was initialized with 101 agents with opinions equal to \textit{strongly disagree} while the remaining agents were respectively assigned with the \textit{disagree} opinion (20 agents) and the \textit{mildly disagree} one (19 agents).

\subsubsection*{Opinion dynamics}
\begin{figure}
    \centering
    \includegraphics[width=\textwidth]{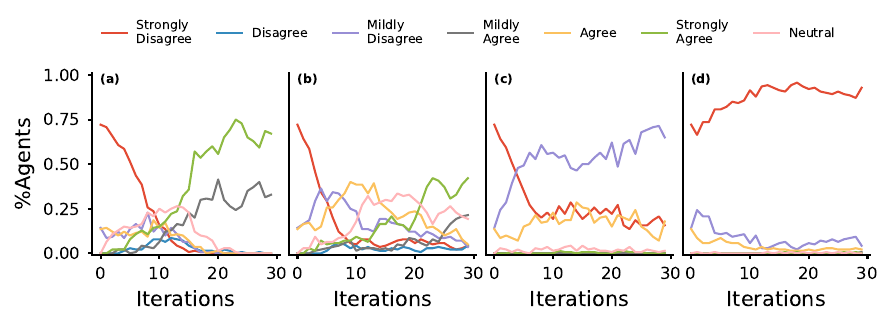}
    \caption{\textbf{Unbalanced scenario - Opinion trends.} Percentages of agents holding each opinion as a function of time, with opinion being one of \textit{strongly disagree} (red), \textit{disagree} (lilac), \textit{mildly disagree} (yellow), \textit{neutral} (pink), \textit{mildly agree} (green), \textit{agree} (grey) and \textit{strongly agree} (blue). Panel (a) and (b) refer to Mistral agents discussing the positive (a) or negative (b) statement. Panel (c)-(d) refer to Llama-agents.}
    \label{fig:unbalanced_trends}
\end{figure}

Following the same structure as in previous paragraphs, we will discuss opinion evolution trends in this scenario.
From Figure \ref{fig:unbalanced_trends} emerges that -- differently from the previous scenarios -- no generalizable trend can be identified across different \acrshort{llm}s. 
In the case of Mistral agents, as shown in Figure \ref{fig:unbalanced_trends}(a) and (b), in either case (same boat or different boat), the \textit{strongly disagree} opinion immediately declines.
In contrast, the other two disagreeing opinions survive in the population. 

At some point, disagreeing agents start becoming neutral and shift toward the \textit{mildly agree} opinion and less toward the \textit{agree} opinion. 
As we can see from the brown and red lines in Figure \ref{fig:unbalanced_trends}(a), after a few interactions, all the population agrees around the statement \say{the boat is the same}, even if with different strengths of agreement. 
If we change the statement to \say{the boat is different}, we can see that the dynamics are very similar, with the \textit{strongly disagree} stance quickly becoming a minority in the population and all the other opinions coexisting for several iterations; however, there is still a tendency towards agreement, with \textit{mildly agree} being the most popular opinion. 
If we consider Llama agents, as we can see from Figure \ref{fig:unbalanced_trends}(c)-(d), not only are the dynamics different from Mistral agents populations but switching how the statement is presented (\say{same} or \say{different}) remarkably impacts the dynamics. 
When the statement is presented in a \say{positive} way, i.e., \say{the boat is the same} (c), we can see that the number of agents holding the \textit{strongly disagree} opinion starts to drop immediately. 
However, the number of agents holding the \textit{disagree} opinion and the number of agents holding the \textit{mildly disagree} opinion increases. 
After a few iterations, there is a majority of \textit{disagreeing} agents (green line) and a minority of \textit{strongly or mildly disagreeing} agents (blue and purple line). 
Moreover, the dynamics are unstable, with the number of agents holding a given opinion increasing and decreasing quickly. 
\textit{Agreeing} agents never manage to emerge in the discussion, and \textit{neutral} agents always remain a minority under 10\% of the population. 
Changing the statement from positive to negative (\say{the boat is different}) changes this pattern we have seen so far, where \textit{strongly disagreeing} agents changed their minds towards other opinions very quickly. 
Instead, the prevalence of \textit{strongly agreeing} agents in the population grows while the two minor initial clusters of \textit{disagreeing} agents become even smaller. 
Also, in this case, \textit{neutral} and \textit{agreeing} opinions never emerge in the population. 

\begin{tcolorbox}[colback=gray!5!white,colframe=gray!75!black,title=Highlight]
Even with unbalanced initial conditions, Mistral agents quickly converge towards agreement, showing more instability when discussing \say{the boat is different}. Llama agents, however, form clusters of disagreement when presented with the positive form, with a majority leaning towards \textit{disagree}. 
In contrast, the negative form results in a substantial majority of \textit{strongly disagree}, with other opinions fading. 
\end{tcolorbox}

Additional figures on the opinion prevalence trends in the polarized scenario can be found in the Supplementary Information (see Supplementary Figures S5-S6). 

\smallskip
\begin{figure}
    \centering
    \includegraphics[width=\textwidth]{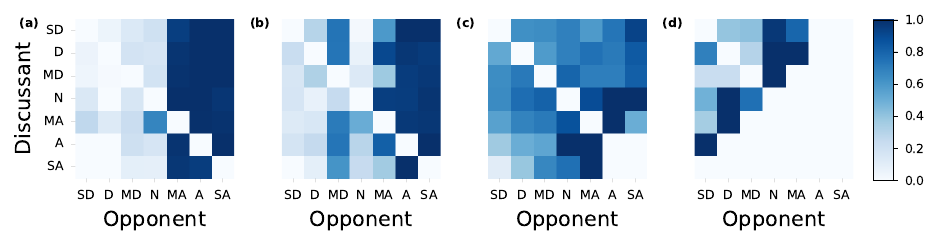}
    \caption{\textbf{Unbalanced scenario - Acceptance rates.} Matrices (a) and (b) represent the \textit{Discussant}'s acceptance rate given both the \textit{Discussant}'s and the \textit{Opponent}'s stance for Mistral agents discussing the positive statement (a) and the negative statement (b); Similarly, (c) and (d) represent results for Llama-agents.}
    \label{fig:unbalanced_matrices}
\end{figure}

These trends may be explained in the light of the acceptance rate matrices in Figure \ref{fig:unbalanced_matrices}. 
As we can see, in the case of Mistral agents, this rapid drop in the number of agents holding a \textit{strongly disagree} opinion is due to a high (close to 1) acceptance rate of \textit{agreeing} opinions, while a lower acceptance rate of \textit{disagreeing} and \textit{neutral} opinions. 
This holds for both the \say{same} and \say{different} scenarios, while in the \say{different} one, the probability of accepting \textit{disagreeing} opinions slightly increases and that of accepting \textit{agreeing} opinion slightly decreases. 
In the case of Llama agents, we need to distinguish between the two cases as the behaviour of the agents is very different. 
In the \say{same boat} scenario, we can see a situation similar to those we saw in previous sections for the balanced and polarized setup: there is a high acceptance rate, with \textit{disagreeing} and \textit{neutral} agents accepting more agreeing opinions while agreeing agents have a more nuanced behaviour with highest values still in the neutral or agreeing section (Figure \ref{fig:unbalanced_matrices}(c)). 
In this scenario, the reject rate is generally very low, except presented statement (Figure \ref{fig:unbalanced_matrices}(d)), we can see that instead, the acceptance rate of \textit{agreeing} opinions is almost always zero and that \textit{disagreeing} agents mainly accept \textit{neutral} positions, while \textit{agreeing} and \textit{neutral} agents mainly accept \textit{disagreeing} positions.
As we can see from the lowest row of the matrix, \textit{strongly agreeing} agents have an almost zero acceptance rate, but because they are not even present in the population.

\begin{tcolorbox}[colback=gray!5!white,colframe=gray!75!black,title=Highlight]
Mistral agents have a stronger tendency towards agreement than acceptance, accepting at higher rates agreeing opinions and at lower rates disagreeing ones. Llama agents have a stronger tendency towards acceptance than agreement when presented with the positive form of the statement; when presented with the negative form, there is a lower tendency to both accept and agree, leading to a majority of strongly disagreeing agents.
\end{tcolorbox}

Additional matrices can be found in the Supplementary Information (see Supplementary Figures S11-S12).

\subsubsection*{Logical fallacies analysis} 
Llama agents were more prone than Mistral agents to adopt fallacious reasonings. 

\begin{figure}[!h]
    \centering\includegraphics[width=\linewidth]{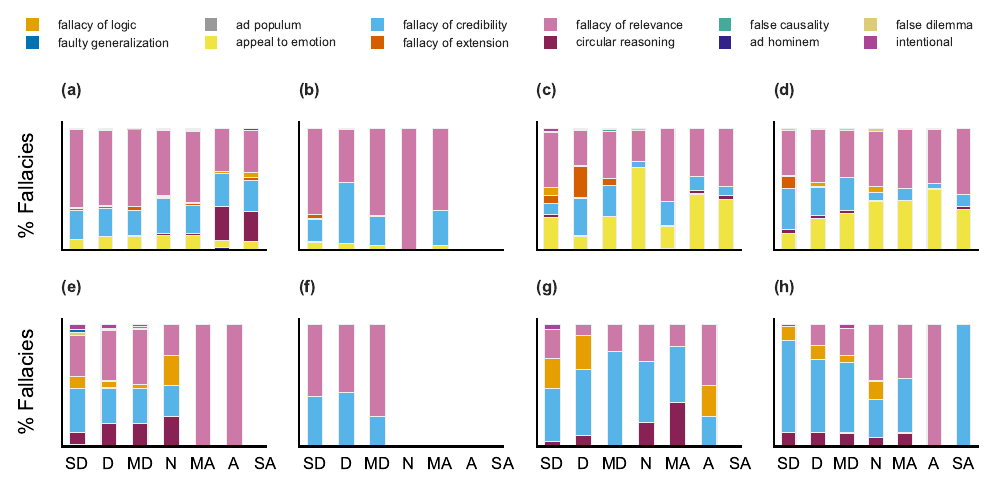}
    \caption{\textbf{Logical fallacies frequency by opinion in the unbalanced scenario.} Percentage of logical fallacies by opinion.
    Panels (a)-(b) refers to Llama \textit{Opponents}, panels (c)-(d) to Mistral. Panel (a) and (c) describe the ``same boat", while (b) and (d) describe the ``different boat" scenario.  
    Figures below refer to \textit{Discussant} for Llama (e)-(f) and Mistral (g)-(h).
    Fallacies of relevance (pink), credibility (light blue), appeal to emotion (yellow), and circular reasoning (plum) are the most common overall.}
    \label{fig:fallaciesUnbalanced}
\end{figure}

Llama \textit{Opponents} exhibit a distribution of logical fallacies similar to that observed in the balanced and polarized scenarios.
This is particularly evident in the discussion of the \say{same boat} statement (Figure \ref{fig:fallaciesUnbalanced}(a)), where \textit{negative, neutral and mildly agreeing} agents frequently rely on fallacies of relevance even with a starting unbalanced distribution.
On the contrary, the analysis of fallacies in the case of the \say{different boat} statement aligns with the findings in the opinion evolution. 
As is evident from Figure \ref{fig:fallaciesUnbalanced}(b), \textit{Opponents} failed to generate more positive opinions, with outcomes reaching only the \textit{mildly agree} level. 
Fallacies of relevance were the most common, with neutral agents relying exclusively on this type of fallacy.

As illustrated in Figure \ref{fig:fallaciesUnbalanced}(c)-(d), Mistral agents exhibit an even more pronounced tendency toward appealing to emotion to exert a persuasive power with respect to Mistral agents in the polarized scenario.  
This tendency is particularly evident in the case of \textit{neutral} agents in the \say{same boat} scenario (Figure \ref{fig:fallaciesUnbalanced}(c)), where fallacies of relevance dominate, with 70\% of statements characterized by this type of reasoning.
In the case of the discussion of the \say{different boat} statement, as in Figure \ref{fig:fallaciesUnbalanced}(d), agents employed a reduced number of statements appealing to emotions compared to the polarized scenario.

\begin{tcolorbox}[colback=gray!5!white,colframe=gray!75!black,title=Highlight]
The unbalanced scenario impacts Mistral with an increase in the production of persuasive utterances targeting the adversary's emotions, while Llama's agents continue to appeal to fallacies of relevance.
\end{tcolorbox}

In contrast, the results for the \textit{Discussants} are more similar to those obtained in the balanced scenario, as the statements are characterized by a limited variety of logical fallacies (Figures \ref{fig:fallaciesUnbalanced}(e)-(h)), especially fallacies of relevance and credibility.

Llama \textit{Discussants} (Figure \ref{fig:fallaciesUnbalanced} (e)-(f)), in the \say{same boat} scenario (Figure \ref{fig:fallaciesUnbalanced}(e)), exhibited circular reasoning in addition to the two fallacies of relevance and credibility. 
A particular case is shown in Figure \ref{fig:fallaciesUnbalanced}(e), as Llama \textit{Discussants} with \textit{mildly agree} and \textit{agree} opinions are characterized by only fallacies of relevance, while \textit{strongly agreeing} agents do not appear in the simulation.

Conversely, in the \say{different boat} discussion (Figure \ref{fig:fallaciesUnbalanced}(f)), \textit{Discussants} never managed to reach \textit{neutral} or \textit{positive} stances, leading to the generation of texts flawed by fallacies of relevance and credibility when answering \textit{Opponents}. 
However, similar to the positive framing of the statement, even in the negative one, Llama\textit{Discussants} holding the \textit{agree} and \textit{strongly agree} opinions consistently employed only fallacies of relevance, whereas agreeing agents were solely characterized by fallacies of credibility.

Shifting to Mistral (Figures \ref{fig:fallaciesUnbalanced}(g)-(h)), agents manifest more evident trends in the usage of flawed utterances. 
In both the positive and negative framing of the statement, agents primarily produced fallacies of credibility with a minor component of circular reasoning and fallacies of relevance when answering the \textit{Opponents}.
Furthermore, even Mistral \textit{Discussants} exhibit a trend similar to one of Llama agents in the \say{different boat} scenario (Figure \ref{fig:fallaciesUnbalanced}(h)). 
Agents holding the \textit{agree} and \textit{strongly agree} stance tend to uniformly make statements with only fallacies of relevance, while \textit{agreeing} agents are only characterized by \textit{fallacies of credibility}.

Table \ref{tab:opinion_change_rates} shows the percentage of \textit{Discussants} changing opinion after reading fallacious arguments by the \textit{Opponents}. 
Although fallacies produced a higher rate of opinion changes in the balanced and polarized scenario, the measure is equal to 50\% also in the current one, with the highest value (59\%) registered with Mistral agents discussing the \say{same boat} statement, while the minimum occurs with 
Mistral agents discussing the \say{different boat} statement.
This suggests that a highly negatively polarised setup in LLM-based simulations may encourage more logical and reasoned persuasive statements. 

\begin{table}[]
\centering
\begin{tabular}{|l|ll|ll|ll|}
\hline
                                & \multicolumn{2}{c|}
                                {\textbf{Balanced}} & \multicolumn{2}{c|}{\textbf{Polarized}} & \multicolumn{2}{c|}{\textbf{Unbalanced}} \\  \hline
                                & \multicolumn{1}{l|}{Mistral}  & Llama  & \multicolumn{1}{l|}{Mistral}   & Llama  & \multicolumn{1}{l|}{Mistral}   & Llama   \\ \hline
\multicolumn{1}{|l|}{Same}      & \multicolumn{1}{l|}{0.49}     & 0.72   & \multicolumn{1}{l|}{0.68}      & 0.77   & \multicolumn{1}{l|}{0.59}      & 0.50    \\ \hline
\multicolumn{1}{|l|}{Different} & \multicolumn{1}{l|}{0.45}     & 0.75   & \multicolumn{1}{l|}{0.56}      & 0.76   & \multicolumn{1}{l|}{0.45}      & 0.50    \\ \hline
\end{tabular}

\caption{Opinion change rates after fallacious arguments across the six scenarios using Mistral and Llama-enhanced agents.}
\label{tab:opinion_change_rates}
\end{table}

\section{Methods}\label{sec:methodology}

In the \acrfull{lod} model, we have a population of $N$ agents, where each agent $a$ is an \acrshort{llm} agent, i.e. an instance of a \acrlong{llm}. Agents are enhanced using AutoGen \cite{autogen}: \say{a framework for creating multi-agent AI applications that can act autonomously or work alongside humans}. Specifically, we exploited AutoGen AgentChat's AssistantAgent, a built-in agent that uses a \acrlong{llm} and has the ability to use tools. 
It serves as a foundational agent that can be customized or integrated into multi-agent conversations.

In our model, each \acrshort{llm} agent holds a discrete opinion $x_a \in \{0, \dots, 6\}$ associated (from 0 to 6) with a negative (\textit{strongly disagree, disagree, mildly disagree}), \textit{neutral}, or positive (\textit{mildly agree, agree, strongly agree}) stance on a given statement $s \in S$ around a given topic $\theta \in \mathcal{T}$.
A statement \( s \) can have a \textit{positive valence}, e.g., ``this is true,'' or a \textit{negative valence}, e.g., ``this is not true.'' 

In our study, we chose the Ship of Theseus paradox as the topic, where the statements were phrased as ``the ship is the same'' (positive valence) and ``the ship is different'' (negative valence). 
To formalize this, we define a function \( \pi(s) \) that maps statements to their valence as follows:
\[
\pi(s) =
\begin{cases} 
+1, & \text{if } s \text{ expresses a positive valence (e.g., ``the ship is the same'')} \\ 
-1, & \text{if } s \text{ expresses a negative valence (e.g., ``the ship is different'')}
\end{cases}
\]

At each discrete time step \( t \), a pair of agents \( (a_i, a_j) \) is randomly selected from this network. 
One agent from the pair is assigned the role of \textit{Discussant} (\( D \)) while the other takes on the role of \textit{Opponent} (\( O \)). 

\noindent \textbf{Prompts}\\
\textit{Discussants} $D$ act according to the following prompt.

\begin{tcolorbox}[colback=gray!5!white,colframe=gray!75!black,title=Discussant Prompt]
\begin{small}
\begin{verbatim}
[INST] 
    ### You {Discussant_opinion} on the reasoning conclusion 
    provided as input.
    Task:
    - Listen to the argument of {Opponent.name} on the reasoning
    conclusions and decide if you maintain your opinion 
    or change it.
    
    ### Constraints: 
    - At the end of each interaction declare if you 
        - 'ACCEPT' {Opponent.name} argument;
        - 'REJECT' {Opponent.name} argument; 
        - 'IGNORE' your original opinion.
    Write your response with the following format: 
    \"My original opinion was I {Discussant_opinion} 
    on the reasoning. 
    After reading your argument my conclusions are: 
    I <ACCEPT|REJECT|IGNORE> your stance because <argument>\"
[/INST]"
\end{verbatim}
\end{small}
\end{tcolorbox}

The role of the \textit{Opponent} is instead modeled by the following prompt: 

\begin{tcolorbox}[colback=gray!5!white,colframe=gray!75!black,title=Opponent Prompt]
\begin{small}
\begin{verbatim}
[INST] 
You {Opponent_opinion} on the reasoning conclusion provided as input. 
Support your opinion by providing personal arguments. 
Avoid using already generated arguments. 

IF {Discussant.name} writes REJECT in his answer, write a second 
statement where you declare if you <ACCEPT|REJECT|IGNORE> his stance. 
Otherwise, conclude the conversation writing a message with 
a single word 'END'. 

### Constraints: 
    - In your first statement you must adhere to your opinion 
    ('{Opponent_opinion}')
    - Write your first response with as: \"I {Opponent_opinion} on the
    provided reasoning conclusions. I think that <argument>\} 
[/INST]",
\end{verbatim}
\end{small}
\end{tcolorbox}

The selected Discussant $a_D$ engages in a discussion with the Opponent $a_O$ on a predefined topic $\theta \in \mathcal{T}$, with the goal of influencing the other's opinion. 
During this interaction, the Discussant $a_D$ and the Opponent $a_O$ are prompted to maintain their initial opinions unless convinced by the argumentation of the other. 

The discussion is started by $a_D$, who asks agent $a_O$ to express their opinion on statement $s$ around topic $\theta$ with valence $\pi(s)$.

In our study we have two different statements:  

\begin{tcolorbox}[colback=gray!5!white,colframe=gray!75!black,title= Positive valence statement {$\pi(s)$ = +1}]
    Theseus set sail to reclaim the throne as king of Athens. During the journey, parts of Theseus's ship began to break or decay; Theseus and his crew replaced these parts as they sailed. Eventually, each part of the ship is replaced. In the end the Ship of Theseus is still the same ship on which he originally sailed.
\end{tcolorbox}

and 

\begin{tcolorbox}[colback=gray!5!white,colframe=gray!75!black,title= Negative valence statement {$\pi(s)$ = -1}, enhanced, 
    breakable,
    skin first=enhanced,
    skin middle=enhanced,
    skin last=enhanced,]
Theseus set sail to reclaim the throne as king of Athens. During the journey, parts of Theseus's ship began to break or decay; Theseus and his crew replaced these parts as they sailed. Eventually, each part of the ship is replaced. In the end, the Ship of Theseus is completely different from the one he originally sailed.
\end{tcolorbox}

The question has the following structure:

\begin{tcolorbox}[colback=gray!5!white,colframe=gray!75!black,title=Discussion initialization, enhanced, 
    breakable,
    skin first=enhanced,
    skin middle=enhanced,
    skin last=enhanced,]
\begin{verbatim}
    What do you think of the following statement?: {s}
\end{verbatim}
\end{tcolorbox}

The \textit{Opponent} is asked to produce a persuasive utterance in response to the \textit{Discussant}, based on their current opinion, with the aim of persuading the \textit{Discussant} and shifting their stance. The \textit{Discussant} then processes the \textit{Opponent}'s response and generates a comment about that statement, expressing whether it was convinced by the \textit{Opponent} or not. The interaction may result in a positive (+1) or negative (-1) change in the \textit{Discussant}'s opinion, or no change (0). Finally, the \textit{Opponent} closes the discussion in one of two ways: if the \textit{Discussant} chooses not to accept the persuasive statement, then it generates a new statement commenting on the current stance of the \textit{Discussant} and thanking it for the discussion. 
This comment does not affect the opinions' status, it simply ends the iteration round.
Otherwise, if the \textit{Discussant} is persuaded by the \textit{Opponent}, the \textit{Opponent} can simply end the iteration with an END keyword.
In the present study we set the number of iterations to $T =$ 30. At each iteration $t$ there are $N$ pairwise random interactions $(a_D, a_O)$. 
\\ \ \\
\noindent \textbf{Logical Fallacies detection}\\
The presence of logical fallacies in text is usually identified through transformers-based models, so the task is often approached as a multi-label classification problem.
In this work, we employ the \texttt{distilbert-base-fallacy-classification} model \cite{HFmodel} obtained from HuggingFace. 
We chose this specific model as it is trained on the dataset used 
by Jin et al. \cite{Jin2022}, which introduces the task of logical fallacies detection and the LOGIC dataset for fallacies.
In the present article, we refer to the 13 fallacies illustrated in the original article by Jin et al. \cite{Jin2022}.

In the following, we present and discuss only the most common fallacies identified in our analysis; readers are referred to the original paper for a full summary.
\begin{itemize}
 \setlength\itemsep{0.1em}
 \item \textbf{Fallacy of credibility}: it consists of an appeal to a form of ethics or authority; 
 \item \textbf{Fallacy of relevance}: the argument relies on premises that are irrelevant to the conclusion. In \cite{copi2014introduction}, it is suggested that premises might be \textit{psychologically relevant} but not \textit{logically relevant}, resulting in an argument that seem apparently correct and persuasive;
 \item \textbf{Appeal to emotion}: this fallacious argument assumes that premises are not relevant to conclusions, but the premises are used as a means to convey a specific emotion aiming to manipulate the beliefs of the reader; 
 \item \textbf{Circular reasoning} (\textit{circulus in probando}): a fallacy characterized by a circularity in the reasonings so that the premises depend on the conclusions and vice-versa. 
\end{itemize}

\section{Discussion}\label{sec:discussion}
This study builds on recent literature \cite{chuang2023simulating, park2024generativeagentsimulations1000, breum2023persuasive} and introduces a \acrfull{lod} to explore how language and social influence drive opinion evolution, with a particular focus on the role of logical fallacies. 
In the studied model, each agent holds a discrete opinion ranging from \textit{strongly disagree} to \textit{strongly agree}. 
In each step of our simulations, a \textit{Discussant} asks an \textit{Opponent} for their opinion on a given topic, to which the \textit{Opponent} responds, attempting to persuade the \textit{Discussant}.
The \textit{Discussant} may then adjust their opinion by ±1 or maintain it, with the process continuing until opinions stabilize or a stopping condition is met. 
In the present work, three initial conditions were studied: (a) balanced, with opinions uniformly distributed; (b) polarized, with only extreme opinions; and (c) unbalanced, with the majority of agents holding extremely negative opinions. 
As a discussion topic, we chose the paradox of the ship Theseus.

For each condition, the discussion statement was framed either positively (\say{The boat is the same}) or negatively (\say{The boat is different}), resulting in a total of six scenarios overall, simulated with two different \acrshort{llm}. 
Our analysis shows a clear and consistent pattern: \acrshort{llm} agents show a strong predisposition to (i) agree with the presented statement and (ii) agree with each other, regardless of initial conditions, statement framing, or the language model used. 
Negative and neutral opinions quickly disappeared in all simulations, resulting in prevailing agreement, either as consensus or majority opinion in favour of the statement.
Mistral agents exhibit a selective acceptance pattern, with a high probability of accepting agreeing opinions and a strong resistance to disagreeing opinions (with the original statement). 
This focused dynamic leads to rapid and decisive convergence towards agreement, although it often results in unstable fluctuations within the spectrum of agreement. 
Llama agents, on the other hand, show a broader openness to accept different, even opposing, views. 
Although this \say{open-mindedness} delays the convergence process, it still leads to agreement over time.

Interestingly, the framing of the statement influences not the direction but the convergence path. 
Positive framing tends to promote smoother transitions and higher acceptance rates for agreeable opinions, while negative framing leads to greater instability, with clusters of agreeable opinions persisting longer before finally converging. 

Despite these nuances, the overall trend remains clear: \acrshort{llm} agents are inherently agreeable, moving consistently towards alignment with the presented statement.
This tendency reflects the design of the language models themselves, which prioritize coherence and alignment in the dialogue and shape the agents' interactions to favour agreement \cite{taubenfeld2024systematic, OviedoTrespalacios2023risks} on shared answers or positions.
This inclination toward agreement may indicate sycophantic tendencies in LLM agents—a propensity to excessively agree with others, sometimes at the expense of logical coherence or accuracy. 
In human-AI interactions, such behaviour can reinforce existing beliefs, foster echo chambers, and exacerbate polarization \cite{deng2024deconstructingethicslargelanguage}.
In our simulations, LLM agents consistently displayed higher acceptance rates than rejection of opposing views, promoting consensus even from initially polarizing conditions. 
In contrast, Mistral agents show a more selective pattern of acceptance. 
Their acceptance is more pronounced when confronted with opinions that agree with the proposed statement. 

In addition to the previously discussed patterns of sycophancy and agreement bias, our study confirms previously investigated patterns in how LLMs generate and respond to persuasive arguments. 
The linguistic analysis shows that (i) LLM agents produce logically flawed arguments in an attempt to persuade others, and (ii) they are influenced by arguments containing logical fallacies, i.e. they change their minds as a result. 
This finding is consistent with the \cite{Payandeh2023, li2024reason, mouchel2024logical} literature: false causality and faulty generalisation fallacies have been consistently documented in various models, including Llama-2 and Mistral \cite{mouchel2024logical}, and GPT-3.5 was actually more likely to be persuaded by fallacious arguments compared to logically sound reasoning \cite{Payandeh2023}. 

In our simulations \textit{Opponents}, tasked with persuading the \textit{Discussants}, frequently employed fallacies to sway their counterparts, which were highly influenced by these fallacious statements and, in the end, produced fallacious utterances themselves. 
Among the most common fallacies produced by \acrshort{llm} agents were those of relevance and credibility, differently from \cite{Payandeh2023}, in which false dilemmas and straw man fallacies were the most common in GPT-3.5 and GPT-4.

Llama agents were more successful in persuading \textit{Discussants}, with approximately 73\% of them changing their opinion (in the \say{same boat} case) after a relevance or credibility fallacies. 
In comparison, Mistral agents were less successful, with only 45-49\% of \textit{Discussants} changing their minds depending on the scenario. 
Llama \textit{Discussants} often resorted to circular reasoning. 
While Mistral \textit{Discussants} were less prone to circular reasoning, they often relied on fallacies of credibility, particularly when starting from a polarized initial condition to strengthen their arguments. 

\section{Conclusion}\label{sec:conclusion}
This study introduces a \acrfull{lod}, allowing for the exploration of how language and social influence shape opinion dynamics. 
By utilizing LLM agents, the study demonstrates how synthetic agents, when left unprompted, tend to agree with a given statement and with other interacting partners and use logical fallacies in their attempts at persuasion and argumentation.

One key limitation of the current framework is the simplicity of the agents. 
In this model, agents are equipped with verbal reasoning skills but lack distinct personalities or cognitive diversity. 
The introduction of more complex agent types - such as those with different decision-making styles, biases or psychological traits - could better replicate the diversity of human interactions \cite{lacava2024openmodelsclosedminds, huang2024emotionallynumbempatheticevaluating}. 

Future extensions of the framework could also benefit from a deeper integration of cognitive biases \cite{chuang-etal-2024-simulating} and demographic factors \cite{wang2023humanoid}, as these elements are known to influence opinion dynamics in the real world. 
Furthermore, the model currently assumes a mean-field scenario, which neglects the structure of real-world social networks. 
Incorporating network features such as clustering, assortativity, or echo chambers could significantly increase the realism of the simulations and improve their ability to replicate polarization dynamics \cite{wang2024decoding, piao2025emergence, zheng2024simulating}.
Preliminary tests with alternative network topologies and more sophisticated opinion dynamics algorithms suggest the potential to capture more complex patterns of interaction.

The exploration of fallacious reasoning in social simulations of LLM agents and its role in opinion dynamics has been approached at a preliminary level in this study, leaving substantial opportunities for future investigation.
The role of fallacies poses challenges not only in the context of social simulations - where agents could potentially be optimised through better prompting, enhanced memory, or other refinements to mitigate fallacious reasoning - but also in human-LLM interactions. 
If LLMs are easily swayed by illogical arguments and tend to validate human perspectives, they may inadvertently reinforce false or potentially harmful beliefs.

To improve the understanding of these dynamics, several directions for future research can be pursued.
One key focus is investigating methods to reduce fallacious reasoning in LLMs, such as through improved prompting, enhanced memory mechanisms, or adjustments to biases. 
Understanding the interplay between memory, bias, and opinion evolution is also critical for analyzing the role of persuasive language in opinion change.
Comparing LLM-based simulations with real-world data from online interactions or controlled experiments can help evaluate (i) the robustness of the framework, (ii) its ability to replicate human behavior, and (iii) the effects of linguistic features on opinion change under controlled conditions.

To summarize, despite its limitations, the framework provides a valuable tool for studying the mechanisms of consensus-building and argumentation in a controlled environment.
The framework could serve as a foundation for exploring the drivers of opinion dynamics and their implications for phenomena such as polarization, bias, and misinformation.

\printglossary[title=List of Abbreviations,type=\acronymtype]

\section*{Supplementary information}

The present article has accompanying Supplementary Information files with figures and tables complementary to those presented in the main text. 



\section*{Declarations}


\subsection*{Availability of data and materials} 
The datasets generated and analyzed during the current study is within this paper or publicly available at \cite{repoproject}.

\subsection*{Code availability} 
Code to replicate simulations and analysis is publicly available at \cite{repoproject}. 

\subsection*{Competing interests} 
The authors declare that they have no competing interests.

\subsection*{Funding} This project is supported by SoBigData.it which receives funding from the European Union—NextGenerationEU—National Recovery and Resilience Plan (Piano Nazionale di Ripresa e Resilienza, PNRR)—Project: “SoBigData.it—Strengthening the Italian RI for Social Mining and Big Data Analytics”—Prot. IR0000013—Avviso n. 3264 del 28/12/2021 (to VP and RG); this work is also supported by the scheme ’INFRAIA-01-2018-2019: Research and Innovation action’, Grant Agreement No 871042 ’SoBigData++: European Integrated Infrastructure for Social Mining and Big Data Analytics’ (to RG); finally this work is supported by: the EU NextGenerationEU programme under the funding schemes PNRR-PE-AI FAIR (Future Artificial Intelligence Research) (to EC and RG)

\subsection*{Author contribution.} EC analyzed the data and wrote the paper. 
VP analyzed the data and wrote the paper. 
GR designed the experiments, performed the experiments and supervised the project. 
DP supervised the project.
All authors read and approved the final manuscript. 

\subsection*{Acknowledgements.} We thank Daniele Atzeni for the valuable feedback and Giuliano Cornacchia for the help in designing plots and figures. 

\noindent

\bibliography{sn-bibliography}

\end{document}